\documentclass[twocolumn,prb]{revtex4}

\usepackage{graphicx,amsmath}

\def\be{\begin{equation}}
\def\ee{\end{equation}}
\def\bea{\begin{eqnarray}}
\def\eea{\end{eqnarray}}
\def\ba{\begin{array}}
\def\ea{\end{array}}

\begin{document}

\title{Compensation driven superconductor-insulator transition}

\author{M. M\"uller}

\affiliation{Department of Theoretical Physics, University of
Geneva, Switzerland}

\author{B. I. Shklovskii}

\affiliation{Theoretical Physics Institute, University of
Minnesota, Minneapolis, Minnesota 55455}

\date{\today}

\begin{abstract}
The superconductor-insulator transition in the presence of strong
compensation of dopants was recently realized in La doped YBCO.
The compensation of acceptors by donors makes it possible to
change independently the concentration of holes $n$ and the total
concentration of charged impurities $N$. We propose a theory of
the superconductor-insulator phase diagram in the ($N$, $n$)
plane. It exhibits interesting new features in the case of strong
coupling superconductivity, where Cooper pairs are compact,
non-overlapping bosons. For compact Cooper pairs the transition
occurs at a significantly higher density than in the case of
spatially overlapping pairs. We establish the
superconductor-insulator phase diagram by studying how the
potential of randomly positioned charged impurities is screened by
holes or by strongly bound Cooper pairs, both in isotropic and
layered superconductors. In the resulting self-consistent
potential the carriers are either delocalized or localized, which
corresponds to the superconducting or insulating phase,
respectively.
\end{abstract}

\maketitle

\section{Introduction}
\label{sec_In}

The superconductor-insulator (SI) transition remains a challenging
and controversial subject after more than two
decades~\cite{Fisher,Finkelshtein,Hebard,Goldman,MaLee,Nandini,Mason,Gantmakher,Shahar,Meir,Baturina,Kapitulnik,FeigelmanIoffeKravtsov}.
In the low temperature limit one can drive the SI transition by
changing the film thickness, the magnetic field, or the concentration of electrons in
gated devices. In high $T_c$ superconductors such as YBCO one can
tune the concentration of holes by changing the oxygen doping.
This leads to a SI transition at small hole concentrations of
about 6$\%$ per Cu site in the CuO plane. From this perspective,
YBCO is essentially a heavily doped semiconductor. It is well
known that upon decreasing the doping a semiconductor undergoes
the metal-insulator transition when the three-dimensional
concentration of dopants $N$ crosses the threshold $Na^3 \approx
0.02$. Here $a =\hbar^{2}\kappa/ m e^2$ is the effective Bohr
radius, $\kappa$ is the dielectric constant, $m$ is the effective
mass and $e$ is the proton charge. In under-doped high $T_c$
superconductors the conducting phase is a superconductor, and one
expects a superconductor-insulator transition at a similar
threshold concentration of dopants.

In semiconductors one can vary the concentration of carriers and
impurities independently using compensation. For example, in a
$p$-type semiconductor doped with $N_A$ monovalent donors
compensation means addition of a concentration $N_D < N_A $ of
donors, so that the concentration of remaining holes $n = N_A -
N_D$ becomes much smaller than the total concentration of charged
impurities $N = N_{A}+ N_{D}$. The metal-insulator transition in
the ($N$, $n$)-plane of a compensated semiconductor was studied
long ago. It was shown~\cite{SE,SEbook} that in heavily doped
samples with $Na^3 \gg 1$ the transition takes place when $n(N)
\sim N/(Na^3)^{1/3}$, as was later verified by experiments (cf., Fig.
13.3 in Ref.~\onlinecite{SEbook}).

Recently~\cite{Ando} it was demonstrated that YBCO crystals can
also be strongly compensated by doping with La. Although many of
the La$^{3+}$ ions substitute for Y$^{3+}$ and are therefore not
electrically active, some La$^{3+}$ ions substitute for Ba$^{2+}$
and hence play the role of monovalent donors compensating oxygen
acceptors. It was shown that the sample of $\rm
{Y_{1-z}La_{z}(Ba_{1-x}La_{x})_{2}Cu_{3}O_{y}}$ with $x = 0.13$
and z = 0.62 is completely compensated at $y = 6.32$, and becomes
$n$-type at $y < 6.32$. Thus, also in high $T_c$ superconductors
the concentration of impurities and holes can be varied
independently. Resistance measurements~\cite{Ando} showed that the
SI transition point non-trivially depends on both $x$ and $y$. In
strongly compensated samples it occurs at much larger
concentration of holes than in standard uncompensated samples.
However, the full phase diagram of the zero-temperature SI
transition in the plane ($N$, $n$) has not been established yet
experimentally. In this paper we predict it theoretically.

\subsection{Global phase diagram}
\begin{figure}[b]
\centerline{\includegraphics[width=0.4\textwidth]{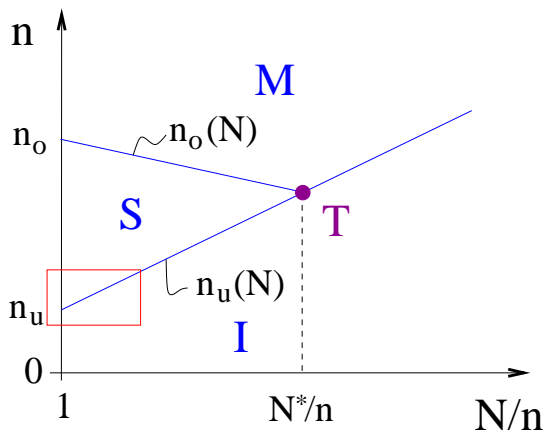}}
\caption{ \label{Fig:GlobalDiagram} Global phase diagram of
compensated high $T_c$ superconductors in the plane $(n, N)$. Here
M, S and I stand for metal, superconductor and insulator,
respectively. The line $n_u(N)$ separates the insulator form a
superconductor, which eventually turns into a metal upon
overdoping beyond $n_o(N)$. We conjecture the existence of a
tricritical point $T=\{n_u(N^*),N^*\}$, where $n_u(N^*)=n_o(N^*)$
beyond which the insulator turns directly into a metal. A large
part of the paper is focused on the lower left corner of the line
$n_u(N)$ which exhibits features of the BEC-to-BCS crossover of
the interacting gas of Cooper pairs in strong coupling
superconductors. }
\end{figure}

Let us start by discussing the gross features of the phase diagram
which are expected, e.g., in compensated high $T_c$ materials such
as La doped YBCO, see Fig.~\ref{Fig:GlobalDiagram}. In the
uncompensated material with $n=N$, we expect a transition from the
insulator to a superconductor at a critical doping $n_u$ (on the
underdoped side) as discussed above. The pairing mechanism is
believed to be at least in part due to spin fluctuations which
become significantly weaker upon exceeding an optimal doping
level. Finally, superconductivity is essentially destroyed on the
overdoped side $(n>n_o)$, or at least $T_c$ is strongly
suppressed. Upon adding the temperature axis to the phase diagram
this leads to the well-known superconducting dome in high
temperature superconductors. As disorder is increased by
compensation (increasing $N/n$), the doping concentration $n_u(N)$
where delocalized states first appear, increases as well. On the
other hand, we expect that the upper critical density $n_o(N)$
decreases because usually disorder diminishes the effectiveness of
the superconductive attraction, while it enhances the competing
Coulomb repulsion. We thus propose that at some compensation $N/n
= N^*/n$ where $n_u(N^*)=n_o(N^*)=n^*$ there may exist a
tricritical point beyond which a direct transition from a
localized insulator to a metal without intermediate
superconducting state takes place. Note that the effect of
compensation is similar to that of a strong magnetic field: both
suppress superconductivity.

In this paper we are not concerned with the transition to a metal
at high doping, nor with the vicinity of the tentative tricritical
point $T=\{n_u(N^*),N^*\}$ in Fig.~\ref{Fig:GlobalDiagram}, where
metal, insulator and superconductor meet. Instead we analyze the
dependence of $n_u$ on the degree of compensation. In the case of
strong coupling superconductivity the latter exhibits interesting
new features in the regime of low densities, reflecting the
crossover from a BEC (Bose-Einstein condensate) to a BCS
superconductor in the interacting gas of preformed Cooper pairs.

\subsection{BEC-to-BCS crossover in the SI transition}
A first attempt to predict the low density part of the SI phase
diagram~\cite{Boris} was based on the toy model of an isotropic
compensated $p$-type semiconductor with a strong (unspecified)
pair-forming mechanism. The size of hole pairs $\xi$ was taken as
a free parameter as determined by a strong coupling mechanism. For
the major part of this paper we will adopt this approach as well.
For simplicity we assume $\xi$ to be independent of the density of
carriers and the disorder, at least in the dilute BEC part of the
phase diagram. However, it would not be difficult to account for
such a dependence (in certain strong coupling models for preformed
pairs one expects such a dependence even in the dilute
regime~\cite{FeigelmanIoffeKravtsov}). In the following we thus
concentrate on the two independent variables $n$ and $N$, taking
$\xi$ as a fixed parameter.

In Ref.~\onlinecite{Boris} two limiting cases of the SI transition
were identified: In the limit of large pairs which overlap
significantly in space, $n\xi^3 \gg 1$, one obtains the standard
Bardeen-Cooper-Schrieffer (BCS) instability of the fermion system.
If disorder is weak the electrons are delocalized and form a dirty
BCS superconductor. This happens essentially at the same critical
density as the metal-insulator transition in a
semiconductor~\cite{SE,SEbook} without superconductivity,
\begin{equation}
\label{nI3}
n=n_1(N) = \frac{N}{(Na^3)^{1/3}}.
\end{equation}
Note that there is no dependence on $\xi$ in (\ref{nI3}), because
electrons are only weakly bound and,
therefore, screen the random potential of charged impurities like
free ones. Here and in all formulae below we omit numerical
coefficients and adopt the scaling approach. The scaling is
controlled by the large dimensionless parameter $Na^3\gg 1$ and
the dimensionless ratio $a/\xi$.

The opposite limit of very small and strongly bound pairs is more
unusual. Upon decreasing the concentration of holes $n$ the SI
transition occurs due to the localization of hole pairs (composite
bosons) in a random potential~\cite{Fisher}. At small external
disorder the bosons undergo a Bose-Einstein condensation (BEC)
while at large disorder the condensate is fragmented and turns
into a Bose insulator (also referred to as Bose glass). This limit
is reached when the pairs are dilute, $n\xi^3 \ll 1$, and can be
considered
as a gas of point-like charged bosons. A similar picture applies
to the case of neutral bosons~\cite{BorisCold,Nattermann}.

As we will rederive below, for point like bosons the border
between the superconducting
BEC phase and the Bose insulator occurs at the hole density%
\begin{equation}
\label{nIII3}
n = n_3(N) = \frac{N}{(Na^3)^{1/5}},
\end{equation}
where typical screened Coulomb wells loose their bound quantum
levels. Of course, the length $\xi$ is again irrelevant, because
the pairs are considered as point-like bosons. Notice that for a
heavily doped system with $Na^{3} \gg 1$, we have $n_{3}(N) \gg
n_{1}(N)$. In other words, in a given disorder, a system of small
pairs delocalizes
at a much higher density $n$ than a system of weakly bound
electrons with larger pair size $\xi$. This reflects the fact
that, at equal density $n$, bosons have less kinetic energy, and
thus one needs more of them to induce their collective
delocalization.

\begin{figure}
\centerline{\includegraphics[width=0.4 \textwidth]{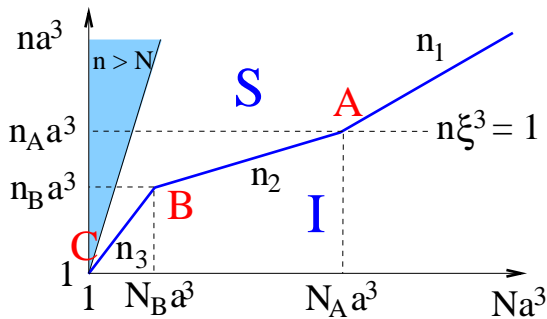}}
\setlength{\columnwidth}{3.2in} \caption{SI-phase diagram in an
isotropic 3D system, on a log-log plot (S stands for
superconductor, I for insulator). The dilute boson part of the
curve (C-B) is described by $n_{3}$, Eq.~(\ref{nIII3}), the
interacting boson part (B-A) by $n_{2}$, Eq.~(\ref{nII3}), and the
standard strong coupling BCS transition beyond A by $n_1$,
Eq.~(\ref{nI3}). The BEC part of the transition line (C-A) only
exists if the pairs are very small, $\xi<a$. In the case of larger
pairs ($\xi > a$), and for weak coupling in general, the  $n_1$
line extends all the way to the point C. The shaded region
corresponds to $n>N$ which is unphysical.} \label{fig:SIT3D}
\end{figure}

The crossover between the above two limiting cases is quite
subtle. In Ref.~\onlinecite{Boris} it was incorrectly assumed that
the limits described by $n_{1}(N)$ and $n_{3}(N)$ require only the
inequalities $n\xi^{3}\gg 1$ or $n\xi^{3}\ll 1$ to be valid, and
hold all the way up to the BEC-BCS crossover line $n\xi^3=1$.
However, this argument neglected the repulsion between bosons as
they become denser, and thus it lead to the incorrect conclusion
that the BEC-BCS crossover line forms a substantial part of the SI
transition line. Below we reconsider this crossover in detail.

This paper contains two main new results. First, in
Sec.~\ref{sec:Is} we show that
while $n_{1}(N)$ is valid all the way up to the BEC-BCS crossover
line $n\xi^3=1$, the BEC part of the SI transition is to a large
extent dominated by an intermediate segment $n=n_2(N)$ of the
transition line at which the chemical potential of repulsive
compact bosons becomes of the order of the amplitude of the
screened Coulomb potential. This segment interpolates between the
above discussed limits $n_1(N)$ and $n_3(N)$, see
Fig.~\ref{fig:SIT3D}. We will see that the BEC regime, $n_2(N)$
and $n_3(N)$, occurs only when the pairs are smaller than the
effective Bohr radius, $\xi< a$. We show below that in this case
transport in the insulating phase is due to the hopping of hole
pairs. In App.~\ref{app:bipolaron} we will discuss bipolarons as
an example of strong coupling superconductivity which can give
rise to such small pairs.

Second, in Sec.~\ref{sec:Ans} we apply similar ideas to a generic
strongly anisotropic superconductors with a layered structure,
such as formed by the CuO or FeAs planes in high $T_c$
superconductors. We arrive at a qualitatively similar phase
diagram in the plane ($N$, $n$) for this case as well, see
Fig.~\ref{fig:SIT2D}. The details of the phase diagram are found
to depend on the ratio between Bohr radius and interlayer
distance.

\section{SI phase diagram of an isotropic superconductor}
\label{sec:Is}
Let us now recall the derivation of the limiting
critical concentrations $n_{1}(N)$ and $n_{3}(N)$. We consider the
case of heavily doped materials, $Na^3\gg 1$, which provides a
large parameter that makes the scaling analysis well controlled.

\subsection{BCS segment of the superconductor-insulator transition line}
We start from the BCS side at high density (large pairs). Let us
divide the sample into cubes of linear size $R$. Due to spatial
fluctuations of the concentrations of donors and acceptors each
cube contains a random impurity charge of arbitrary sign and with
an absolute value of the order of $e(NR^3)^{1/2}$. At the scale
$R$ such randomly fluctuating charges create a random potential
energy relief of amplitude
\begin{equation}
eV(R)\sim \frac{e^{2} (NR^3)^{1/2}}{\kappa R} =  \frac{e^{2}
(NR)^{1/2}}{\kappa}. \label{VR}
\end{equation}
This energy diverges at large $R$, so that screening even by a
small concentration of holes $n$ is crucial. To discuss this
screening we estimate the characteristic fluctuating density
$\delta N(R)$ of impurity charges at the scale
scale $R$:
\begin{equation}
\delta N(R) = \frac{(NR^3)^{1/2}}{R^3}=
\left(\frac{N}{R^{3}}\right)^{1/2}. \label{DN}
\end{equation}
The concentration $n$ of carriers can be redistributed between
wells and hills of the random potential. This redistribution
screens all the scales $R$ for which $\delta N(R) \leq n$ or,
in other words, for $R \geq R_s$, where
\begin{equation}
R_s = \left(\frac{N}{n^{2}}\right)^{1/3} \label{RS}
\end{equation}
is the nonlinear screening radius~\cite{SE,SEbook}. All scales $R
< R_s$ remain unscreened, because even when all electrons are
transferred from all the hills of the potential energy to all its
wells they are not able to level off the charge density of such
fluctuations. Since $V(R) \propto R^{1/2}$ among remaining scales
the most important contribution to the random potential is given
by $R=R_s$. Thus, the amplitude of the nonlinearly screened random
potential energy is
\begin{equation}
eV(R_s)=\frac{e^{2}}{\kappa}\frac{N^{2/3}}{n^{1/3}}. \label{V}
\end{equation}
So far we have dealt only with the electrostatic energy of holes
and neglected their kinetic energy. At $T=0$ all kinetic energy is
of quantum origin, and we should find the conditions under which
it is small enough so that the above described picture of
localized electrons is valid. Clearly the potential energy
Eq.~(\ref{V}) is able to localize electrons with concentration $n$
if it is larger than the Fermi energy of holes in its wells
$\epsilon_{F}(n) = \hbar^2 n^{2/3}/2m$  ($m$ being the effective
mass). In the opposite case $\epsilon_{F}(n) \gg eV(R_s)$ the
Fermi sea covers the typical maxima of the potential energy relief
and the semiconductor behaves like a good conductor, see
Fig.~\ref{fig:droplets}. Equating $eV(R_s)$ and $\epsilon_{F}(n)$
we obtain the critical concentration $n_{1}(N)$ for the SI
transition, as given by Eq.~(\ref{nI3})~\cite{SE,SEbook}. Note
that the non-linear screening theory requires $nR_s^3=N/n\gg 1$
which is always fulfilled in strongly compensated materials.

In the delocalized phase electron screening becomes linear, the
screening radius being given by the standard Thomas-Fermi
expression \bea \label{TF} r_s= \left(\frac{1}{e^{2}}
\frac{d\mu}{dn}\right)^{1/2} =   \frac{a}{(na^3)^{1/6}}, \eea and
the amplitude of the screened potential relief equals $eV(r_s) =
e^{2}(Nr_s)^{1/2}/\kappa$. As expected, at the transition $n=n_1$,
these two quantities match the corresponding expressions
Eq.~(\ref{RS}) and Eq.~(\ref{V}) pertaining to the insulating
side.

\subsection{BEC segments of the superconductor-insulator transition line}
In the above discussion the notion of strong pairing attraction
and preformed pairs was irrelevant. However, as we follow the
transition line Eq.~(\ref{nI3}) to lower densities, we may finally
reach the crossover to the BEC regime, which takes place when
strongly bound hole pairs become dilute, i.e., when $n_1\xi^3=
(Na^3)^{2/3}(\xi/a)^3=1$. This corresponds to the point $A$ in
Fig.~\ref{fig:SIT3D} and the densities
\bea
n_A^{(3d)}a^3 &=& (a/\xi)^3,\\
N_A^{(3d)}a^3&=& (a/\xi)^{9/2}.
\eea

Under the assumption of heavy doping, $Na^3\gg 1$, the crossover
to the BEC regime can only happen when the pair size is much
smaller than the Bohr radius, $\xi < a (Na^3)^{-2/9} < a$. For the
sequel we will assume that the pairs are very small $\xi < a$.
Since we will be using the concept of strongly bound pairs a lot,
we briefly recall the essential elements of strong coupling
superconductivity.

\subsubsection{Strong coupling superconductivity}
The physics of a fermion gas subject to attractive interactions
(but in the absence of disorder) has been studied in detail in
Refs.~\onlinecite{Eagles,Leggett,NS}, and is now a very active
field of studies in the context of cold atoms~\cite{PhysRep}. The
authors of Ref.~\onlinecite{NS} considered electrons with a
mutually attractive potential of size $V_{k,k'}\sim V$ for $k\leq
k_0$, and rapidly decaying for larger $k$. If the interaction
potential between two holes is too weak to produce a bound state
($V< V_c \sim 1/mk_0$), the fermions are essentially unbound, and
only an exponentially narrow range of energies around the Fermi
level participates in pairing, the gap being of the order of
\bea \label{BCS} \Delta\approx \frac{4 k_F^2}{m} \exp[
-1/\nu(E_F)V_{k_F,k_F}],
\eea
where $\nu(E_F)$ is the density of
states at the Fermi level. On the other hand, if the mutual
interaction is strong, bound states of two single carriers exist,
and at low density the fermions organize into preformed pairs with
a typical size $\xi\approx [(V/V_c-1)k_0]^{-1}$ and a pairing
energy $E_{\rm pair}\sim \hbar^2/m\xi^2$. As long as the pairs are
dilute, $n\xi^3\ll 1$, the chemical potential for the addition of
pairs, $\mu$, and the gap function $\Delta$ are much smaller than
the pairing energy, \bea \label{mu_dilute}
\mu &\sim & E_{\rm pair}\, (n\xi^3) \sim \frac{\hbar^2}{m}n\xi,\\
\label{Delta_dilute} \Delta &\sim & E_{\rm pair}\, (n\xi^3)^{1/2}.
\eea
However, when the pairs become dense,  the pair chemical
potential is dominated by the Fermi energy of its constituting
fermions,
\bea
\label{mu_dense} \mu\simeq E_F = \frac{\hbar^2
n^{2/3}}{m}.
\eea
At the same time the gap function $\Delta$
crosses over to its strong coupling BCS form, i.e.,
Eq.~(\ref{BCS}) with an exponent of order~\cite{footnote1} $O(1)$,
\bea
\label{Delta_dense} \Delta\sim E_F.
\eea
In this dense regime the pairing energy is dominated by the gap
function $E_{\rm pair}\approx \Delta > \hbar^2/m\xi^2$.

\subsubsection{Very dilute bosons}
In order to derive the critical concentration $n_{3}(N)$ of Cooper
pairs (charge $2e$ bosons) at the SI transition, we notice that
the above calculation of the nonlinear screening radius $R_s$
(\ref{RS}) and the random potential energy created by screened
charged impurities (\ref{V}) remains unaltered in the scaling
sense.

The difference between the gas of composite bosons and that of
weakly bound fermions lies in their quantum kinetic
energy~\cite{Boris}. Due to the weak effect of Pauli's principle
on strongly bound Cooper pairs, a large number of them can occupy
a given localized level of a potential well, keeping the quantum
kinetic energy low. Therefore, the condition of delocalization of
Cooper pairs is much more stringent than the condition $eV(R_s) <
\epsilon_{F}(n)$ which applies to fermions. A sufficient condition
for the delocalization of a compact Cooper pair is that a typical
well of the random potential does not contain any localized level,
or $eV(R_s) < \hbar^{2}/mR_{s}^2$, where $m$ is the effective mass
of pairs, which we assume to be of the same order as that of
electrons. This condition is also {\em necessary} if mutual
repulsions can be neglected, as we will discuss below. Solving the
equation
\begin{equation}
eV(R_s) = \frac{\hbar^{2}}{mR_{s}^2}, \label{BE}
\end{equation}
for $n$ and using Eqs. (\ref{RS}) and (\ref{V}) we find the
critical concentration of the SI transition given in
Eq.~(\ref{nIII3}). This derivation clearly demonstrates why
$n_{3}(N) \gg n_{1}(N)$. According to Eq.~(\ref{V}) the potential
energy amplitude $eV(R_s)$ decreases with increasing $n$. To
achieve delocalization it has to be pushed below the quantum
kinetic energy of the clean system. This requires larger $n$ in
the boson case and thus leads to $n_{3}(N) \gg n_{1}(N)$.

\subsubsection{Moderately dilute, interacting bosons}

So far, following Ref.~\onlinecite{Boris}, we have taken into
account all the Coulomb interactions. However, we have neglected
the short range repulsive interaction between composite bosons.
Such a repulsion is related to the Fermi nature of individual
holes, which becomes important if two pairs of holes overlap
within their length $\xi$. This short range interaction can be
described by the well-known expression for the chemical potential
$\mu(n)$ of a non-ideal gas of bosons of concentration $n$ with a
scattering length $\xi$:
\begin{equation}
\mu = (\hbar^2/m)n\xi, \label{mu}
\end{equation}
which is also confirmed by the result (\ref{mu_dilute}) for dilute
systems of strong coupling superconductivity. This chemical
potential reflects the extra quantum kinetic energy due to the
mutual repulsion of the pairs. Note that it matches the Fermi
energy $E_F=\hbar^2 n^{2/3}/m$ when the BEC-BCS crossover
$n\xi^3=1$ is reached.

The delocalization criterion (\ref{BE}) discussed above remains
relevant as long as the density is low enough such that $\mu$ is
smaller than the typical localization energy $\hbar^2/mR_s^2$.
However, at an impurity density $N=N_B$, the chemical potential of
the critical insulator ($n=n_3(N)$) becomes of the order of the
typical amplitude of the random Coulomb potential Eq.~(\ref{V}).
This marks the point $B$ in Fig.~\ref{fig:SIT3D}, beyond which the
delocalization is driven by the mutual repulsion between bosons.
The crossover in the transition line occurs at the densities
\bea
n_B^{(3d)}a^3&=& (a/\xi)^2,\\
N_B^{(3d)}a^3&=& (a/\xi)^{5/2}.
\eea
On the low density side the $n_3(N)$ line ends at point
$C$ which corresponds to the uncompensated limit $n_3 a^3= Na^3=
1$. At higher densities, $N>N_B$ we need to compare the quantum
kinetic energy Eq.~(\ref{mu}) to the amplitude of potential
fluctuations, Eq.~(\ref{V}), similarly as in the BCS regime. This
leads to the new segment of the transition line \bea \label{nII3} n=n_2(N)=
\frac{N^{1/2}}{(a\xi)^{3/4}}, \eea which interpolates between
points $A$ and $B$ in Fig.~\ref{fig:SIT3D}.

We can confirm this result by calculating the linear screening
radius $r_s$ in the delocalized Bose gas. Using Eq.~(\ref{mu}) to
compute the compressibility we find the Thomas-Fermi screening
radius from Eq.~(\ref{TF}) as $r_s = (a\xi)^{1/2}$. One easily
verifies that this
linear screening radius matches the non-linear screening radius
(\ref{RS}) at the transition line (\ref{nII3}). Similarly, one can
check that along the line $n_1$ the linear screening radius of the
conducting side matches the non-linear screening radius $R_s$ on
the insulating side. The linear screening in the very dilute
superfluid above the line $n_3$ is found~\cite{Boris} to be \bea
\label{rs} r_s = \left(\frac{a}{n}\right)^{1/4}, \eea which again
matches $R_s$ at the transition line $n_3$.

The fact that the SI transition line undergoes a kink at the
BCS-BEC crossover is very similar to the case of the
superfluid-insulator phase transition in a neutral gas of
attractive fermions~\cite{BorisCold}. However, at lower density
$n_2$ lies below the BCS-BEC crossover line, contrary to what was
claimed in Ref.~\onlinecite{Boris}.

The results obtained so far in this section can be summarized in
the following concise manner: The chemical potential of a gas of
composite bosons of size $\xi$ , localized into a region of linear
size $R$ is given by
\bea \label{mu3d} \mu(n,R)= \max\left[\frac{\hbar^2}{m
R^2},\frac{\hbar^2}{m} n\min(\xi,n^{-1/3})\right].
\eea
The first term of the righthand side refers to the ground state
energy in a well of size $R$. At higher density
$\mu(n,R)$ is dominated by the second one, describing the
interaction energy (\ref{mu}) of repulsive bosons (for
$n\xi^3<1$), and the Fermi energy $E_F\sim n^{2/3}$ of the BCS
regime (for $n\xi^3>1$), respectively.

The SI transition occurs when the chemical potential dominates
over the amplitude of the screened impurity potential, i.e., when
\bea
\mu(n,R_s)\sim eV(R_s).
\eea
This can be reformulated as
\bea
\label{SIT}
\max\left[\left(\frac{n}{n_{3}(N)}\right)^{\frac{5}{3}},\,
\frac{n}{n_{1}(N)} \min[1,(n\xi^3)^{\frac{1}{3}}]\right]=1,
\eea
which defines the transition line in the whole $(n,N)$-plane, as
plotted in Fig.~\ref{fig:SIT3D}.

\subsection{The nature of the insulating regime}

\subsubsection{Droplets in the insulator}

It is important to understand the insulating phase in some more
detail. Deep in the insulator, the charge density $n$ is by no
means homogeneously distributed. Instead the holes or Cooper pairs
fill deep wells, where they form puddles of high density, while
the rest of the space is completely void of carriers.

\begin{figure}[b]
\centerline{\includegraphics[width=0.4 \textwidth]{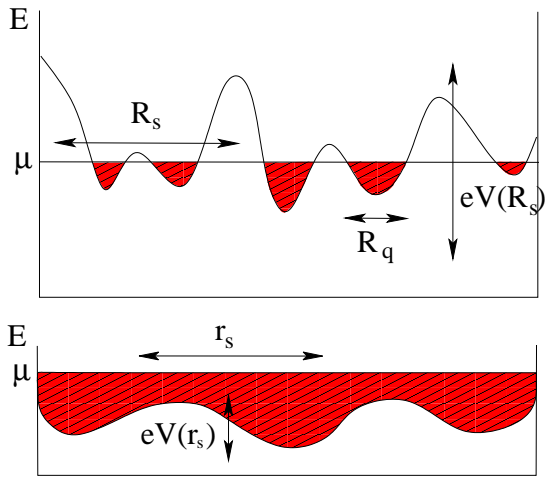}}
\setlength{\columnwidth}{3.2in} \caption{{Top: Droplets in the
insulator. The carriers assemble in non-percolating fragments of
size $R_q < R_s$. Bottom: In the conductor, the wells are not deep
enough to localize the carriers. The latter delocalize due to
their quantum kinetic energy, $\mu>V(r_s)$ ($\mu=E_F$  in the BCS
limit of dense pairs).} \label{fig:droplets} }
\end{figure}

To determine the chemical potential and the size of puddles we can
argue as follows:
when $n\ll n(N)$ only the deep wells of the landscape are
populated with carriers (see Fig.~\ref{fig:droplets}).

Suppose the carriers fill a well of linear size $R < R_s$. Its
typical depth $eV(R)$ is given by Eq.~(\ref{VR}) , and it contains
an excess impurity charge of order $eQ(R)=e(NR^3)^{1/2}$. Upon
filling the well with carriers, their chemical potential raises
continuously with respect to the bottom of the well. Assume that
when the chemical potential reaches $eV(R)$ we have filled in
$Q_{q}(R)$ particles. If $R$ is small, $Q_{q}(R)< Q(R)$, that is,
the exclusion principle or the repulsion between bosons limits the
number of carriers we can fill into the well. In large wells, we
can at most fill in $Q(R)$ particles before turning the well into
a hump. However, those larger wells will not be filled
homogeneously with particles. Rather, they split into smaller
droplets for which $Q_{q}(R)\approx Q(R)$. The latter relation
defines~\cite{SEbook} a typical droplet size $R_q$: \bea
\mu\left(n=n_q\equiv \frac{Q(R_q)}{R_q^3}\right)= eV(R_q). \eea
Here we have to use the expressions for the chemical potential
given above in Eq.~(\ref{mu3d}).

Analyzing the three regimes of the phase diagram (BCS side,
repulsive bosons and very dilute bosons) we find the following
results. When $n\xi^3>1$ ($N > N_A^{(3d)}$ on the transition
line), $\mu$ is given by $E_F(n)$, and we find the droplet
size~\cite{SEbook}
\bea
R_{q,1} = \frac{a}{(a^3N)^{1/9}}.
\eea

In the repulsive boson regime, $ (N\xi^3)^2 < n\xi^3 < 1$ (or
$N_B^{(3d)}<N < N_A^{(3d)}$ on the transition line), we find
instead
\bea
\label{Rq2} R_{q,2}=(a\xi)^{1/2}.
\eea

Eventually, in the lowest density regime, $n\xi^3 < (N \xi^3)^{2}$
where the bosons do not significantly interact, $R_q$ is simply
the typical localization radius in the disorder potential. It is
obtained from $\hbar^2/mR^2 \sim eV(R)$ as
\bea
 R_{q,3}= \frac{a}{(a^3 N)^{1/5}}.
\eea

In all three cases, the insulator consists mostly of puddles of
size $R_q$ which are well separated and do not percolate. One can
verify that the SI transition occurs when the droplets grow to the
size of the non-linear screening radius, $R_q = R_s =
(N/n^2)^{1/3}$. Indeed, at this point droplets of size $R_s$ start
to percolate, which induces the delocalization transition.

Note the remarkable fact that in all insulating regimes the density
of carriers in the above droplets is the same as the critical density
of the corresponding segment of the SI transition, $n_q =
Q(R_q)/R_q^3= n(N)$.

\subsubsection{Level spacing in droplets}

Before we turn to the role which droplets play in the transport
properties of the insulator phase, we have to discuss the level
spacing in a typical droplet. The typical cost to add another
carrier into a droplet is $\delta = R_q^{-3}\, d\mu/dn(n=n_q)$.
This is essentially the level spacing of the considered droplet.
Interestingly this quantity turns out to be equal to the charging
energy $e^2/R_q$. This holds both in the dense BCS-like part and
in the interacting boson regime of the phase diagram. In the very
dilute boson regime, the quantity of interest is not the level
spacing $\delta$, but the typical kinetic energy scale of a
localized wavefunction, which again turns out to be equal to the
charging energy $e^2/R_q$. Thus, for the above (spontaneously
originating droplets) we do not have to distinguish between
one-electron level spacing and charging energy.

This unique energy scale is important in determining whether all
carriers are paired or whether it is energetically favorable to
break up a pair and redistribute the constituting holes onto two
different droplets. The cost of such a break up is the pairing
energy, while the maximal energy gain is of order $\delta \sim
e^{2}/R_q$. Thus the criterion for having all carriers paired up
in the ground state is
\bea
\delta \sim e^2/R_q < E_{\rm pair} .
\eea

In the BEC regime the pairing energy is given by $E_{\rm pair}\sim
\hbar^2/m\xi^2$. In the BCS regime it is even bigger, if the BCS
coupling remains strong, see Eq.~(\ref{Delta_dense}). Thus, if
$\xi < a$, it is never favorable to break Cooper pairs, i.e. the
insulating state is always a Bose glass. Moreover the droplets are
actually superconducting at low enough temperatures. In tunneling
experiments they should show a hard gap with coherence peaks on
its shoulders, despite the absence of global phase coherence among
the droplets.

\subsection{Variable range hopping transport in the insulator}

The above implies that for systems with small pairs, $\xi < a$,
the low temperature transport is Efros-Shklovskii variable range
hopping of Cooper pairs between droplets.
This yields a
conductivity \bea \label{R_ES} \sigma(T)= \sigma_0 \exp\left[-
\left( \frac{T_{\rm ES}}{T}\right)^{1/2}\right]\,, \eea with a
characteristic temperature \bea \label{TES2} T_{\rm ES}= 2.8
\frac{(2e)^2}{\kappa \ell_2}, \eea where $\ell_2$ is the effective
localization length of a Cooper pair. This prediction agrees
qualitatively with experimental data~\cite{Ando,Andoprivate}.

However, there is an exception to the above assertion that pairs
prevail in the insulator if $\xi < a$. Namely, if the
dimensionless BCS coupling $\lambda$ decreases with increasing
density in the BCS regime, the pairing energy $E_{\rm pair}$ can
become exponentially suppressed at high densities. (This
presumably happens on the overdoped side of high $T_c$
superconductors.)

When $E_{\rm pair}\approx \Delta$ falls below the level spacing in
typical droplets, $\delta$, it becomes favorable to break up pairs
and redistribute the carriers on different droplets. One can
verify that at the same time the parity gap (the extra cost for
having an odd number of particles per droplet) becomes smaller
than the level spacing~\cite{MatveevLarkin}. In this situation the
ground state of the system is a Coulomb glass of unpaired
fermions. Consequently, the low temperature transport is again of
the form (\ref{R_ES}), but with a characteristic temperature \bea
T_{\rm ES}= 2.8 \frac{e^2}{\kappa \ell_1}, \eea which is roughly 8
times smaller than (\ref{TES2}), because the localization length
of a hole is about twice as big as that of a pair, $\ell_1 \sim 2
\ell_2$. When pairs are not very strongly bound ($\xi > a$), as
well as in the case of weak coupling, only the BCS segment of the
superconductor-insulator border survives. In the weak coupling
case, it can easily happen that $E_{\rm pair}< \delta$ in the
droplets of the insulator. In this case they contain odd or even
numbers of holes, and the low temperature variable range hopping
is dominated by unpaired holes.

So far we have not specified any particular strong coupling
mechanism which leads to the preformed pairs on the insulating
side of the SI transition. In App.~\ref{app:bipolaron} we discuss
an explicit example of strong coupling superconductivity, which
allows one to formulate a direct microscopic criterion for the
condition $\xi < a$. As discussed above, the latter is necessary
to observe the BEC part of the SI transition in a heavily doped
system.

\subsection{Coulomb correlation energy}

As we saw above the long range Coulomb interactions considered
within mean field approximation play a major role in our theory.
However, the correlation energy produced by the Coulomb
interaction between nearest neighbors has been neglected so far.
We should thus make sure that corrections to the chemical
potential due to Coulomb correlations are subdominant with respect
to the leading term given in Eq.~(\ref{mu3d}). According to
Foldy~\cite{Foldy} the energy per particle in a disorder-free,
Coulomb interacting Bose system in 3d is \bea u_{\rm
Cb}=\frac{e^2}{a}{\rho_s^{-3/4}} = \frac{\hbar^2}{m} n^{2/3}
\rho_s^{5/8}=\frac{\hbar^2}{m} n^{1/4}a^{-5/4}, \eea where
$\rho_s= (a^3 n)^{-1/3}$. One can verify that this quantity is
indeed smaller than $\mu(n, R_s)$, along the whole SI phase
transition line, consisting of the segments $n_1(N)$, $n_2(N)$ and
$n_3(N)$.

\section{SI phase diagram of a layered superconductor}
\label{sec:Ans}

\subsection{General theory}
\subsubsection{Non-linear screening in a layered system}

In this section, we extend our previous arguments to the case of
anisotropic layered superconductors. We assume that with respect
to the motion along the $c$-axis ($z$-axis) all holes reside in
the lowest spatial quantization mode
of the narrow quantum wells defining the conducting $ab$-planes
($x,y$-plane) perpendicular to the $c$-axis. These parallel wells
are located at a distance $d$ from each other, each well
containing holes with the two-dimensional concentration $nd$.
Impurities of both signs are randomly distributed between these
narrow quantum wells. We assume again that a strong attraction
between the holes of a given well leads to preformed pairs
(composite bosons) with a size $\xi$ in the plane of the well. To
simplify things we will first assume an isotropic dielectric
constant $\kappa$. Modifications due to anisotropy will be
discussed in the next subsection.

\begin{figure}[b]
\centerline{\includegraphics[width=0.4
\textwidth]{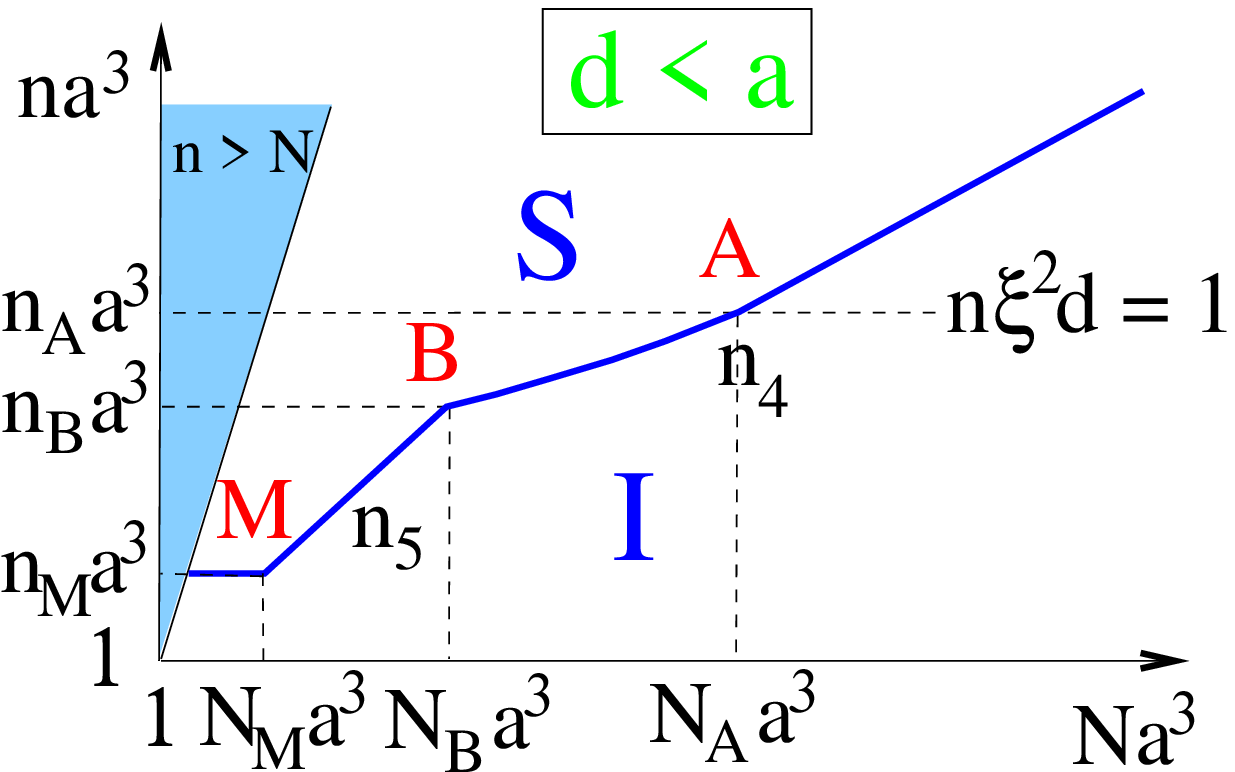}} \setlength{\columnwidth}{3.2in}
 \vspace{.4cm}
\centerline{\includegraphics[width=0.4
\textwidth]{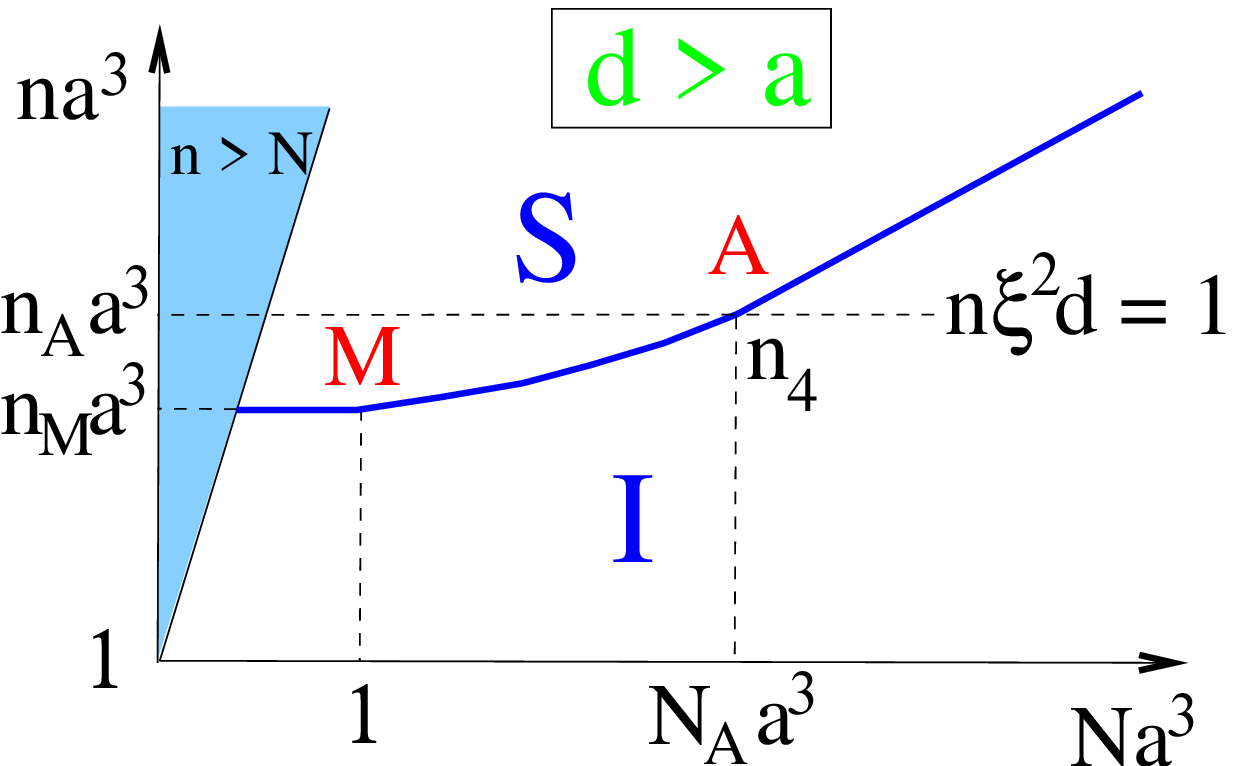}} \setlength{\columnwidth}{3.2in}
\caption{The superconductor-insulator transition in the plane $(n,
N)$ for the model of a layered high $T_c$ superconductor (on a
log-log scale). S stands for superconductor, I for insulator. The
shaded region $n>N$ is unphysical. (a) Narrowly spaced layers,
$d<a$. At lowest densities the SI transition is a Mott transition in
the layers. The very dilute boson part of the curve (M-B) is
described by $n_5$, Eq.~(\ref{nIII2}), while the interacting boson
and BCS parts (B-A and beyond) are described by $n_4$,
Eq.~(\ref{nI2}). (b) Widely spaced layers, $d>a$. The dilute boson
regime does not exist. In both cases, the point $A$ where a weak
BEC-BCS crossover takes place, exists only if $(ad)^{1/2} >\xi$ or
$a>\xi$, respectively.}
\label{fig:SIT2D}
\end{figure}

Let us define again the effective Bohr radius
in the $ab$-plane, $a = \hbar^{2}\kappa/ m e^2 $.
For the insulating phase we need to understand the non-linear
screening in a system containing impurities in the bulk and
screening carriers confined to planes. There
are two limits of this screening problem. When the non-linear
screening radius is bigger than the distance between layers, $R_s
> d$, we can use our results for the isotropic 3d case, cf.
Eqs.~(\ref{RS},\ref{V}). On the other hand if $d > R_s $, the
potential fluctuations within each plane are screened
independently.
For these two cases we have obtained the phase diagrams of
Fig.~\ref{fig:SIT2D}. The specific expressions for the various
lines are derived below.

The non-linear screening radius for the case $d>R_s$ has been
derived in Refs.~\onlinecite{Gergel78,SE86}. Let us cover a
conducting plane by densely packed cubes of linear size $R<d$.
Fluctuations of the charge among these cubes are of the order of
$e(NR^3)^{1/2}$. The random potential they create in the planes
can be screened by redistributing the charge of two-dimensional
holes, $endR^{2}$ between potential hills and wells of linear size
$R$, if the latter is large enough. We find the nonlinear
screening radius $R_s$ by equating $e(NR^3)^{1/2}= endR^{2}$,
which yields \bea R_s= \frac{N}{(nd)^2}. \eea Only scales of the
random potential with $R < R_s$ survive the screening. Thus, the
amplitude of the remaining random potential is \bea \label{V2}
eV_s = \frac{e^2}{\kappa}\frac{N}{nd}. \eea

At small enough hole concentration $n$ the screening radius is
always bigger than $d$. However, upon approaching the SI
transition, either of the above screening scenarios
may apply. As we will see, the first case applies to small
separations of layers, $d < a$, while the screening of independent
layers governs close to the SI transition if $d > a$.

\subsubsection{Narrowly spaced layers, $d < a$}

The difference between layered materials and isotropic ones is due
to the confinement of the carriers to the layers. The quantum
kinetic energy, or the chemical potential of pairs confined to a
region of linear size $X$ in the plane, evaluates to \bea
\label{mu2d} \mu(n,X)= \max\left[\frac{\hbar^2}{m
X^2},\frac{\hbar^2}{m} \frac{n d}{\max\left[1,\log(1/\xi^2 n
d)\right]}\right]. \eea

Note that in 2d there is only a logarithmic difference between the
Fermi energy $E_F$ in the fermion fluid and the interaction energy
per particle in the Bose gas, both scaling essentially as
$(\hbar^2/m) n d$. The logarithmic dependence of $\mu$ on the pair
size $\xi$ in the interacting boson regime is
well-known~\cite{Schick71}. Note that the logarithm is replaced
by unity at the BEC-BCS crossover point $nd\xi^2 =1$. The first
term in (\ref{mu2d}) is the kinetic energy of a single boson or
fermion in a well of size $X$, the relevant size in the context of
non-linear screening being $X=R_s$.

We will show below that the logarithmic effects due to the finite
size of pairs $\xi$ is only relevant for the transition when the
pairs are small, $\xi<(a d)^{1/2}$. Let us thus first discuss the
opposite case $\xi >(a d)^{1/2}$. The chemical potential  for
pairs, Eq.~(\ref{mu2d}), then scales in the same way as that for
fermions without any superconducting correlations. Thus the
results below describe equally well the metal-insulator transition
of unpaired fermions confined to planes.

Delocalization and hence the insulator-conductor transition, takes
place roughly when $\mu(n,R_s)= eV_s$. In the high density regime
where the Fermi energy dominates, the transition occurs at \bea
\label{nI2} n_4= \frac{N^{1/2}}{(ad)^{3/4}}. \eea

The difference between this result and Eq.~(\ref{nI3}) is due to
the confinement of holes to the $ab$-planes. At low densities, the
random potential $eV_s$ competes against the kinetic energy due to
the confinement of carriers to regions of size $R_s$ in the plane.
>From this we find the critical density \bea \label{nIII2}
n_5=\frac{N}{(Na^3)^{1/5}}, \eea which is the same as in the
isotropic case, Eq.~(\ref{nIII3}).

The line (\ref{nIII3}) continues down to densities $n_M=1/a^{2}d$
which is the minimal density required to drive a Mott transition
in the layers. Note that for $d< a$ this minimal density is higher
than in the isotropic case, since here the carriers are confined
to narrowly spaced planes.

The crossover between $n_4(N)$ and $n_5(N)$ occurs at the densities
\bea n_B^{(2d)} a^3 &=& \left(\frac{a}{d}\right)^2
, \quad \quad
N_B^{(2d)} a^3= \left(\frac{a}{d}\right)^{5/2}.
 \eea
The transition lines $n_4$ and $n_5$ can also be derived by
approaching from the conducting side, in close analogy to the
isotropic case.

It is justified to deal with fermions and thus to ignore the
logarithm in Eq.~(\ref{mu2d}) if we are still on the BCS side of
the BCS-BEC crossover, i.e., if $\xi^{2}nd >1$ down to $n=
n_B^{(2d)}$. This condition is equivalent to $\xi > (a d)^{1/2}$
as we anticipated above. In this case, the crossover to the dilute
boson (BEC) regime occurs along the line $n_5$ only,
without affecting the shape of the SI transition line.

Let us now discuss effects which arise if fermions bind into small
pairs of size $\xi<(a d)^{1/2}$.
As in the isotropic case, there is a BCS-BEC crossover at the
point $A$ in the phase diagram, where $n_A^{(2d)}=1/\xi^2d
> n_B^{(2d)}$ and \bea n_A^{(2d)} a^3&=&\frac{a^3}{\xi^2d},
\quad\quad N_A^{(2d)}
a^3=\left(\frac{a}{d}\right)^{1/2}\left(\frac{a}{\xi}\right)^{4}.
\eea However, here the difference between the fermion regime and
the interacting boson regime results only in a logarithmic factor
correcting the line $n_4$ to \bea \label{nI2_2} n_4=
\frac{N^{1/2}}{(a d)^{3/4}} \log^{3/4}\left(\frac{(a
d)^{3/4}}{\xi^2 d N^{1/2}}\right). \eea

\subsubsection{Widely spaced layers, $d > a$}

If the spacing between layers is larger than the Bohr radius we
need to compare the potential fluctuations (\ref{V2}) to the
chemical potential (\ref{mu2d}). As above, we will find that when
pairs are large, $\xi>a$, they do not affect the phase transition
line, which then becomes equivalent to the metal-insulator
transition of an unpaired fermion system. In the high density
regime, equating $E_F$ to $eV_s$ from Eq.~(\ref{V2}), we find the
transition line \bea \label{nI2_larged} n_4=
\frac{N^{1/2}}{a^{1/2}d}. \eea The non-linear screening radius
remains constant $R_s=a$ along the SI transition line.

It turns out that, contrary to the case $d < a$ discussed above,
the line (\ref{nI2_larged}) describes the SI transition down to
dopant densities where $N_M a^3=1$ and $n=n_M=1/a^2d$, while the
first term in (\ref{mu2d}) never becomes relevant. Once the
dopants are dilute, $Na^3<1$, one leaves the regime of heavy
doping. The potential for individual carriers is then dominated by
the closest impurity charge. Under these conditions the
delocalization takes place as a standard Mott transition in the
planes. It occurs when the nearest neighbor distance in the planes
is of order $a$, i.e., when \bea \label{n6} n=n_M=\frac{1}{a^2d}.
\eea Again, it is justified to deal with fermions and to neglect
logarithmic factors if $\xi^2 n d>1$ holds down to  $n=n_M$. This
is equivalent to the condition of large pairs, $\xi>a$.

However, in the case where fermions are strongly bound into pairs
of size $\xi<a$, there are logarithmic corrections to the phase
boundaries. One finds that the line (\ref{nI2_larged}) turns
slightly upward,
\bea \label{nI2_larged_2}
n_4=\frac{N^{1/2}}{a^{1/2}d} \log^{1/2}\left(\frac{a^{1/2}d}{\xi^2 d
N^{1/2}}\right),
\eea
for $n_M< n<n_A^{(2d)}=1/\xi^2d$. The full phase diagram is
shown in Fig.~\ref{fig:SIT2D}b.

\subsection{Anisotropic dielectric constant}
Here we refine the above analysis and take into account the
anisotropy of the dielectric constant in a layered system. We use
$\kappa_z$ for $\kappa_{zz}$ and $\kappa_x$ for
$\kappa_{xx}=\kappa_{yy}$. We also define the average dielectric
constant as $\kappa = (\kappa_{x}^{2}\kappa_z)^{1/3}$.

In order to derive the SI transition line in the presence of an
anisotropic dielectric constant we first switch to the new
coordinate frame $(x', y', z')$, where $x' = x/\kappa_{x}^{1/2}$,
$y' = y/\kappa_{x}^{1/2}$ and $z' = x/\kappa_{z}^{1/2}$. In this
frame~\cite{LL} the Coulomb interaction of a charged impurity with
a hole becomes isotropic $e^2/\kappa^{3/2}r'$. At the same time,
the concentrations $N$ and $n$ are transformed, too: $N' =
\kappa^{3/2} N$, $n' = \kappa^{3/2} n$.

Let us first treat the case of narrowly layered systems. In the
new frame we have similarly to Eqs.~(\ref{RS},\ref{V}) $R'_s =
N'^{1/3}/n'^{2/3} = N^{1/3}/\kappa^{1/2} n^{2/3} $ and $eV(R'_s) =
e^{2}N'^{2/3}/\kappa^{3/2} n'^{1/3} = e^{2}N^{2/3}/\kappa
n^{1/3}$. Thus, returning to the laboratory system we arrive back
at Eq.~(\ref{V}) for the amplitude of the screened potential
$eV_s$, and to the same SI transition line (\ref{nI2}) with
redefined $\kappa = (\kappa_{x}^{2}\kappa_z)^{1/3}$. However, note
that the notion of the nonlinear screening radius $R_s$ becomes
anisotropic. Characteristic potential wells have a scale $X_s =
\kappa_x^{1/2}R_s'= (N^{1/3}/n^{2/3})\alpha^{1/6}$ in the $(x,y)$
plane, where $\alpha = \kappa_{x}/\kappa_z$. On the other hand,
the scale perpendicular to the planes is $Z_s = \kappa_z^{1/2}R_s'
%N^{1/3}/\kappa^{1/2} n^{2/3} \kappa_{z}^{1/2}
= (N^{1/3}/n^{2/3})\alpha^{-1/3} = X_s \alpha^{-1/2} < X_s $. This
anisotropy modifies the critical concentration $n_5$ (i.e., the
very dilute boson limit) to \bea \label{nIII2_aniso}
n_5=\frac{\alpha^{1/5}N}{(Na^3)^{1/5}}. \eea It also affects the
criterion on the smallness of pairs that is required for a BEC
regime with logarithmic factors to exist. The criterion follows
from $n_B^{(2d)}\xi^2d<1$ where $n_B^{(2d)}$ is the crossing point
of $n_4$ and $n_5$. This yields the requirement
\bea
\label{xicrit_smalld} \xi<\alpha^{1/6}(ad)^{1/2}, \quad\quad
(d<d^*).
\eea
The crossover from widely to narrowly spaced
layers can be obtained from the criterion $Z_s=d$. It occurs at
the spacing \bea d=d^*=\frac{\kappa_z}{\kappa}a= \alpha^{-2/3}a.
\eea

In the widely spaced case one finds
\bea eV_s=
\frac{e^2}{\kappa^{3/2}}\frac{N'}{n'd'}=\frac{e^2}{\kappa^{3/2}}\kappa_z^{1/2}\frac{N}{nd},
\eea
and thus the corrected transition line
\bea n_4=
\left(\frac{\kappa_z}{\kappa}\right)^{1/4}\frac{N^{1/2}}{a^{1/2}d}
= \alpha^{-1/6}\frac{N^{1/2}}{a^{1/2}d}. \eea

The Mott transition at low density now takes place at a density
\bea n_M=n_B^{(2d)}=\frac{1}{a_x^2d}=\frac{\alpha^{1/3}}{a^2 d},
\eea where $a_x=\alpha^{-1/6}a$ is the effective Bohr radius in
the plane. It is obtained by comparing kinetic and Coulomb energy
in the plane, $\hbar^2/m a_x^2 = e^2/[\kappa^{3/2} a_x']=
e^2/[\kappa^{3/2}(a_x/\kappa_x^{1/2})]$. Logarithmic corrections
occur in this case if $n_M d\xi^2<1$, i.e., for \bea
\label{xicrit_larged} \xi< a\alpha^{-1/6}, \quad\quad (d>d^*).
\eea  Note that the critical lines $n_4$ for narrowly and widely
spaced systems match when $d=d^*$. The same holds for the critical
size of pairs necessary to have a BEC-like regime.

\subsection{Is the BEC limit of this theory applicable to high $T_c$ superconductors?}

Above we have obtained results along two lines. First, for
relatively large pairs (in the BCS regime) we predict the SI
transition line given by Eqs.~(\ref{nI2}) and (\ref{nIII2}).
Second,  for small pairs we have discussed an additional
logarithmic factor originating from BEC effects. One may question
whether the value of $\xi$ in high $T_c$ superconductors is small
enough so that the condition $nd\xi^{2} \leq 1$ for a BEC-like
regime with extra logarithmic factors is realistic. The most
frequently cited number for under-doped uncompensated YBCO for the
superconducting coherence length is $2$~nm. However, for the case
of strong coupling of holes the size of pairs, $\xi$, can be
smaller than the coherence length. ARPES data in YBCO indicate
that actually~\cite{Campuzanoprivate} $\xi \sim 1$nm. Using that
that the boundary of the superconducting dome on the underdoped
side occurs at $(n_u d) = 0.06a^{-2}_0$ where $a_0\sim 0.4$nm is
the lattice constant of the two-dimensional Cu lattice, one finds
that the SI transition at $T=0$ empirically happens when
$nd\xi^{2} \approx 0.4<1$. This means that the BEC part of our
diagram Fig.~\ref{fig:SIT2D} is marginally relevant. In
iron-arsenide superconductors the small value $\xi < 2$~nm was
recently found~\cite{Ong}. Since $n_u d a_0^2$ has a similar value
as in cuprates, this leads to $n_{u}d\xi^2 \sim 1$ in this new
family of superconductors as well.

The low density (BEC) regime might indeed be experimentally
relevant if one adopts a popular interpretation of the pseudogap
which is observed in underdoped
samples~\cite{Randeria,Uemura,Deutscher,Renner,PhysRep,Campuzano}.
The latter assumes that the pseudogap is due to preformed hole
pairs with large binding energy ($E_{\rm pair}> T_c$), the pairs
being localized by disorder at low doping density. If such an
interpretation is correct, the small $N$ part of our diagram
Fig.~\ref{fig:SIT2D} may be relevant for high $T_c$
superconductors.

As we have seen the SI boundary reflects the BEC-BCS crossover in
the form of extra logarithmic factors in the critical density
$n_4$ only if $\xi$ is sufficiently small, i.e., if condition
(\ref{xicrit_smalld}) or (\ref{xicrit_larged}) is satisfied. Let
us discuss this condition for the example of
Bi$_2$Sr$_2$Cu$_2$O$_{6+\delta}$ (Bi-2201). The mean distance
between copper planes is $d=12.3 \AA$, and the lattice spacing in
the planes is $a_0=5.36 \AA$. From recent optical
measurements~\cite{vanHeumen}, one can extract the effective mass
of carriers as $m_{\rm eff}\approx 3-4 m_e$ in the underdoped
regime. The dielectric constant along the c-axis is~\cite{BSCCO}
$\kappa_z=18.9$. We are not aware of direct measurements of
$\kappa_x$, but usually, the anisotropy is relatively
modest~\cite{Tajima}, e.g., $\alpha = \kappa_{x}/\kappa_z = 0.99$
in Nd$_2$CuO$_4$, or $0.7$ in Pr$_2$CuO$_4$. Neglecting the
anisotropy we can use $\kappa=\kappa_z$ and the effective mass to
estimate the Bohr radius in Bi-2201  as
$a\approx a_x \approx a_z\approx 3\AA$. We can compare this to an
alternative estimate obtained as follows: We assume that the SI
transition of uncompensated materials is essentially a Mott
transition of doped carriers, which is known to occur roughly
when~\cite{MottDavis} $(n_ud) a^2= c$ with~\cite{fn2} $c\approx
0.04$. Using an approximate value for $n_ud\approx 0.06a_0^{-2}$
we find $a \approx 4.4 \AA$, in rough agreement with the above
calculation based on the effective mass. This material thus
certainly corresponds to widely spaced layers, $d>a$. The
estimated Bohr radius $a$ is of the same order as the typical pair
size $\xi$ in strongly underdoped samples. The requirement
(\ref{xicrit_larged}) for observing the SI transition in the BEC
regime is thus just marginally satisfied in this standard cuprate
compound.

More favorable conditions for the crossover to the BEC limit may
be expected in materials with high dielectric constants (such as
in La$_{2-x}$Sr$_x$CuO$_4$), which increases the Bohr radius. A
similar tendency can be expected from a small effective mass
(small band mass and/or small mass renormalization), provided it
does not occur simultaneously with an increase of the pair size
$\xi$.

The application of our theory to certain particular
superconductors may require further adjustments of the model. For
example, in $\rm {Y_{1-z}La_{z}(Ba_{1-x}La_{x})_{2}Cu_{3}O_{y}}$
acceptors are divalent and we have to define proper variables for
the phase diagram. We can use $N_A = (y - 6)/v_{\rm uc} $ for the
concentration of divalent oxygen acceptors (excessive oxygen),
$N_D = x/v_{\rm uc}$ for the concentration of  monovalent donors,
and $n = 2N_A - N_D$ for the concentration of holes. Here $v_{\rm
uc}$ is the volume of the unit cell. It is easy to show that the
concentration $N = 4N_A + N_D$ plays the role of the effective
concentration of monovalent charged impurities. Indeed, for
randomly distributed impurities in a given volume $R^3$, the
variances of the donor, acceptor and net charge number
distribution are equal to $N_{A}R^3$, $N_{D}R^3$ and
$(p^{2}N_{A}+N_{D})R^{3}$ respectively, where $p$ represents the
valence of the acceptor. So the effective concentration of
monovalent charged impurities is not $N = N_A + N_D$ but $N =
p^{2}N_{A} + N_{D}$. For excess oxygen atoms one has $p=2$, and
the coefficient 4 in the expression for $N$ reflects the enhanced
role of divalent charge in the creation of potential fluctuations.

There is a further complication for YBCO, in that a fraction of
holes does not reside in CuO planes, but in CuO chains. This
should be taken into account when comparing our theory with YBCO
data. However, most other high $T_c$ layered superconductors do
not suffer from such a complication.

\section{Conclusion}
In conclusion we have established the phase diagram for the
superconductor-insulator transition in heavily doped, strongly
compensated semiconductors endowed with a strong superconductive
coupling mechanism. The phase transition line at large impurity
and carrier density coincides essentially with the well-known
metal-insulator transition in doped semiconductors. However, if
Cooper pairs are tightly bound, such that $\xi< a$, there is a low
density (BEC) regime where preformed pairs are dilute even at the
SI transition. In this regime we have established two new segments
of the SI transition line which reflect that a gas of compact
bosons is more compressible than an equally dense gas of weakly
interacting fermions.

Recently, an interesting system exhibiting a direct SI transition
upon doping has been discovered in the form of boron-doped
diamond~\cite{diamondexp}. The latter can be simultaneously doped
by both donors and acceptors and thus constitutes a promising
system in which one might observe the effects we predict for the
isotropic 3d case. It would also be interesting to test our
predictions numerically, e.g., following the lines of recent work
which investigated the interplay of superconductivity and
localization in strong disorder~\cite{diamond}.

We have extended considerations from the isotropic case to layered
systems such as the cuprates. In this case, we have found new
equations for the SI line which can be verified experimentally. We
showed that due to the smaller phase space for bosons in the
plane, the crossover to the BEC regime manifests itself on the
phase transition line only by an additional logarithmic factor.
Except for the logarithmic factors, our results also apply to the
metal-insulator transition in layered, strongly doped fermion
systems.

Apart from determining the phase transition line, we have
established the properties of the insulating phase. We assert that
in the presence of strong superconducting couplings all fermions
are paired, and hence at lowest temperatures, transport is due to
the variable range hopping of Cooper pairs. This kind of transport
may be observable in compensated diamond and high $T_c$ materials.

We are grateful to Y. Ando, D. Basov, J. T Devreese, A. Kamenev,
B. Spivak, V. Turkowski and D. van der Marel for useful
discussions. The authors acknowledge the hospitality of the Aspen
Center for Theoretical Physics, where part of this work was done.
MM acknowledges support by the Swiss National Fund for Scientific
Research under grants PA002-113151 and PP002-118932.
%\end{theacknowledgments}

\appendix
\section{Small pairs in bipolaron systems}
\label{app:bipolaron}

Fr\"ohlich polarons are quasiparticles arising in systems with
strong electron phonon coupling. An extensive study of polarons is
given in Ref.~\onlinecite{VerbistDevreese}, based on Feynman's path
integral approach, giving accurate results in dimensions $d=2,3$.
If the coupling strength is large enough polarons can bind into
strongly bound pairs which finally undergo a SI transition if
they are dense enough.

An essential ingredient for strong coupling is a
significantly small ratio between the electronic dielectric
constant $\kappa_{\rm el}$, and the static one, $\kappa >\kappa_{\rm el}$,
\bea \eta\equiv \frac{\kappa_{\rm el}}{\kappa}. \eea

Note that the static dielectric constant is the one which enters
the screening problems discussed in the main text.

The electron-phonon coupling is characterized by the coupling constant,
\bea
\alpha&=&\frac{e^2}{2\hbar \omega_{LO}}\left(\frac{2 m \omega_{LO}}{\hbar}\right)^{1/2}\left(\frac{1}{\kappa_{\rm el}}-\frac{1}{\kappa}\right)\\
&\approx& \frac{e^2}{2\hbar \omega_{LO}}\left(\frac{2 m
\omega_{LO}}{\hbar}\right)^{1/2}\frac{1}{\kappa_{\rm el}}=
\frac{\xi}{a_\infty}.
\eea

Here, $m$ is the band mass and $\omega_{LO}$ the long wavelength
optical phonon frequency. In the last step it was assumed that
$\eta\ll 1$. Further,
\bea
a_\infty =
\frac{\hbar^2\kappa_{\rm el}}{me^2}
%\eta \frac{m_{\rm pol}}{m}a
\ll a , \eea is an "effective Bohr radius" built from the
electronic dielectric constant. However, the Bohr radius which
appears in the theory presented in the main text is given by \bea
a=\frac{\hbar^2\kappa}{m_{\rm pol}e^2}, \eea where $m_{\rm pol}>m$
is the polaron mass. At large coupling, one finds~\cite{Verbist92}
$m_{\rm pol}/m\propto \alpha$. Note that $a$ can be much larger
than $a_\infty$ if $\eta \alpha \ll 1$.

At strong coupling polarons can bind into pairs, so-called bipolarons.
The bipolaron radius is usually only $10-20\%$ larger~\cite{VerbistDevreese}
 than the radius of a single polaron, which is given by
\bea
\xi= \left(\frac{\hbar}{m \omega_{LO}}\right)^{1/2}=\alpha a_\infty.
\eea

These pairs are stable if the coupling is sufficiently strong
($\alpha>\alpha_c$), and if the ratio $\eta$ is sufficiently
small. The critical values have been
computed~\cite{VerbistDevreese} to be $\alpha_c=2.9$ (2D) and
$6.8$ (3D).

For not too strong couplings $\alpha>\alpha_c$ the maximal
admissible ratio of dielectric constants which allows for
bipolaron formation was found to be (for both 2d and 3d)
\bea
\eta_c%1-\frac{\sqrt{2}\alpha}{U_c(\alpha)}
%&=&\frac{1.63-\sqrt{2}}{1.63+\sqrt{2}\alpha_c/(\alpha-\alpha_c)}\nn\\
= \frac{(\gamma-1)(\alpha-\alpha_c)}{\gamma(\alpha-\alpha_c)+\alpha_c}.
\eea
with $\gamma\approx 1.63/\sqrt{2}\approx 1.15$. Note that $\eta_c\to 0$ as $\alpha\to \alpha_c$.

We are eventually interested in the possibility of small pairs
with $\xi< a$. Note that even though a large coupling $\alpha$
implies $\xi > a_\infty$, a small ratio $\eta\ll 1$ still allows
one to have
\bea \frac{\xi}{a} = \alpha\,\eta\, \frac{m_{\rm pol}}{m} \sim
\alpha^2\eta < 1. \eea
This is precisely what is needed to observe
the BEC part of the SI transition line.

At very strong coupling, larger conglomerates of polarons can form
stable bound states. The formation of such multipolarons has not
been studied systematically yet, apart from an analysis at
asymptotically  strong coupling~\cite{Devreese93}. For
multipolarons to exist, the coupling needs to be considerably
stronger, $\alpha\gg \alpha_c$. At a given large $\alpha$, the
most stable bound state will depend on the ratio of dielectric
constants, $\eta$. The smaller $\eta$, the more polarons can bind
together. For example, at asymptotically strong coupling one finds
bipolarons for $0.046=\eta^{(3)}_{c}<\eta<\eta^{(2)}_c=0.079$, and
larger multipolarons for $\eta<\eta^{(3)}$.

Multipolarons containing an even number of polarons will be
compact bosons, which eventually undergo a Bose Einstein
condensation in sufficiently weak disorder. Up to numerical
prefactors, the SI-transition for such multipolarons would be of
the same nature as the one discussed in the main text for bosons
formed by pairs of carriers.

%------------------------------------------------------------------------%


\begin{thebibliography}{99}
\bibitem{footnote1} {Usually the BCS coupling constant $\lambda= \nu(E_F)V_{k_F,k_F}$
increases with density, so that for a strong coupling
superconductor one expects (\ref{Delta_dense}) to remain valid at
high density. However, under certain circumstances (e.g., in
SrTiO$_3$, see Ref.~\onlinecite{SrTiO3}), the interaction strength
can decrease more strongly with increasing density than the
density of states at the Fermi level increases. In this case,
$\lambda$ may decrease with increasing density, and the small
exponential $\exp(-1/\lambda)$ in Eq.~(\ref{BCS}) can become
relevant again.}
\bibitem{fn2} The value of $c$ has to be taken
with a grain of salt in the present context. The presence of
divalent acceptors in YBCO, e.g., may enhance the effective
disorder and thus increase $c$. On the other hand, interaction
effects tend to decrease the value of $c$.
\end{thebibliography}
\end{document}